# Improving Deep Learning-based Defect Detection on Window Frames with Image Processing Strategies

Jorge Vásquez, Hemant K. Sharma, Tomotake Furuhata, and Kenji Shimada

*Abstract*—Detecting subtle defects in window frames, such as dents, scratches, and bends, can be challenging and critical for ensuring product quality and reputation. Traditional machine vision systems may not accurately identify unexpected and random defects, particularly in complex environments with varied object orientations and lighting conditions. Moreover, machine learning methods may struggle with changing environmental lighting conditions and small datasets, limiting their effectiveness in defect detection. To overcome these limitations, we introduce InspectNet, a hybrid deep learning vision-based method that employs advanced image processing techniques to improve defect detection accuracy and reduce errors under adverse lighting conditions. InspectNet also reduces the need for labeled data using the correct image processing technique method for each detection, resulting in a more efficient and generalized inspection process. Our experiments show that InspectNet outperforms other machine learning pipelines, achieving an IoU 9.7% higher than the U-net method. These results demonstrate the potential impact of InspectNet on automating the window-frame inspection process on construction sites, providing a reliable and accurate alternative to manual inspection while reducing time and cost. InspectNet represents a significant step towards improving product quality inspection and protecting companies' reputations in the construction industry.

## I. INTRODUCTION

THE construction industry is currently confronted with the formidable challenge of meeting the escalating demand for quality inspection while contending with a decreasing number of inspectors available. In order to tackle this issue effectively, there exists an urgent requirement for automated robotic inspection systems to target specific objects. Human vision, with its inherent limitations, can often overlook minor and subtle defects that necessitate a certain level of expertise or experience on the part of the inspector. By addressing this need, automated robotic inspection holds tremendous potential for improving the overall quality assessment process. One such object is aluminum window frames, susceptible to surface defects such as scratches, dents, and bends that can significantly impact product quality and safety. Detecting these defects during and after installation ensures product quality and prevents costly faults. However, these defects can be difficult to detect due to complex lighting conditions and variations in colors and materials. This situation highlights the need for more accurate defect detection methods to ensure quality window frame inspection.

In recent years, machine and deep learning methods have proven effective in detecting defects in many industrial cases [6], [9], [30], [35], [41]. However, these methods often struggle with small prepossessing datasets, interpreting ambiguous inspector or defect criteria, and handling extrinsic factors such as illumination. Specifically, complex lighting conditions can produce shadow occlusions, contrast, and color distortion, making it challenging to accurately identify and locate surface defects. Image Processing Techniques (IPTs) have been developed to enhance image attributes [23], [3]. However, these techniques have limitations when detecting unexpected and random defects, mainly out of the laboratory environments [4], [2], [22]. Therefore, there is a need for further research and development to overcome these challenges and improve the accuracy of automated defect detection systems combining deep learning-based methods with image-enhanced IPTs. By addressing these issues, manufacturers can ensure the quality of their products while reducing the costs associated with manual inspection methods.

In order to tackle the challenges arising from adverse lighting conditions in the detection of defects on window frames, this paper presents InspectNet. InspectNet is a hybrid deep learning approach that combines the power of deep learning with a carefully selected strategy of image processing techniques, as illustrated in Fig. 1. The aim of InspectNet is to enhance the accuracy and effectiveness of surface defect detection in window frames.

This work did not receive any funding. *(Corresponding author: Jorge Vásquez).*

Jorge Vásquez is with the Department of Mechanical Engineering, Carnegie Mellon University, Pittsburgh, PA 15213, USA (e-mail: jivasque@andrew.cmu.edu).
Hemant K. Sharma is with Department of Mechanical Engineering, Carnegie Mellon University, Pittsburgh, PA 15213, USA (e-mail: hemantks@andrew.cmu.edu).
Tomotake Furuhata is with the Department of Mechanical Engineering, Carnegie Mellon University, Pittsburgh, PA 15213, USA (e-mail: tomotake@cmu.edu).
Kenji Shimada is with the Department of Mechanical Engineering, Carnegie Mellon University, Pittsburgh, PA 15213, USA (e-mail: shimada@cmu.edu).

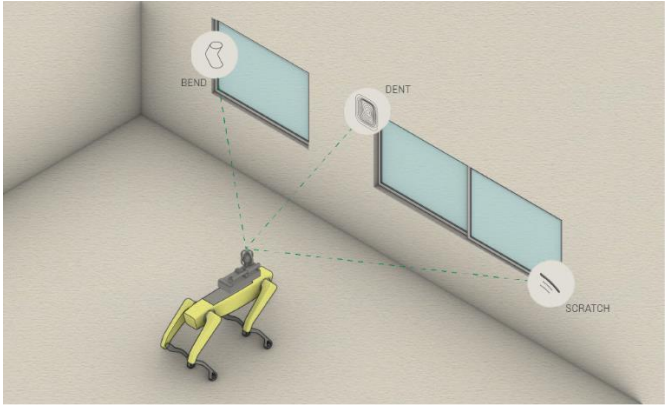

**Fig. 1.** Visualization of our Robotic Inspection of Window Frames.

Our proposed method aims to remove shadows, neutralize intensity and contrast, and accurately locate damage in window frames by combining a fast, lightweight deep learning algorithm with an ensemble illumination enhancement strategy. Our method is specifically tailored to enhance the performance of defect detection methods used in mobile machine-vision systems, especially in complex lighting environments that pose a challenge to such systems. Our approach is tested for mobile robotic platforms, making it highly versatile and adaptable for various environments. We demonstrate the effectiveness of our approach in real-world conditions using four different datasets for experimentation, where detecting window frame defects can be challenging. This is particularly significant given the increasing automation of building inspection systems. Overall, InspectNet represents a significant step forward in defect detection for delicate objects such as window frames. InspectNet is a powerful tool for enhancing the accuracy and efficiency of automated building inspection systems.

The main contributions of this work are:

- **IPT Impact Analysis:** An IPT impact evaluation tool that enables organizations to make data-driven decisions on IPT strategies to improve defect detection accuracy in visual inspection tasks.
- **Hybrid Deep Learning Detection Method:** Our paper introduces a hybrid Deep Learning method that achieves a robust mean IoU of 92% when applied to RGB images, demonstrating its high accuracy and suitability for various applications.

This paper is structured as follows: Section II provides an overview of related work in the field. Section III presents the proposed method, and Section IV details the experiments performed and reports the results obtained. Lastly, Section VI concludes the paper, highlighting the essential findings and contributions of the study.

## II. RELATED WORK

The related work section of this paper explores two primary categories of machine vision systems employed for surface defect detection tasks: traditional approaches and machine learning-based methods. Traditional approaches utilize computer vision models that rely on image processing techniques and statistical analysis to detect surface defects. However, these methods face challenges when confronted with unexpected and randomly located defects, particularly in dynamic scenarios featuring complex lighting conditions. In contrast, modern systems leverage machine learning-based methods that employ neural networks for high-quality surface inspection. These methods harness the power of neural networks to detect and classify defects. Nevertheless, they also encounter challenges stemming from uncertainties related to inspector specifications and changes in defect requirements, particularly in dynamic environments characterized by complex lighting conditions. The following section provides a comprehensive overview of these approaches, highlighting their strengths and limitations.

### A. Traditional Machine Vision methods

Machine vision techniques have been critical in detecting defects for several years. These methods apply various algorithms and statistical models to detect surface damage. The traditional approach to machine vision involves using algorithms to analyze a dataset. Among the most widely used techniques for automatic defect detection using conventional methods is Automatic Threshold [21] for different applications in glass defects [5], in textile manufacturing [33] and other applications. Entropy and image dynamic thresholding for road crack segmentation [23]. Edge Detection using Retinx Algorithm for the surface defect in linear CCD [24], this technique has also been used for other applications as pavements defects [40], and for flexible Integrated Circuit Packaging substrates [14]. Regarding statistical methods, morphological processing has been used for defect detection using genetic algorithms [42], and Fourier Series has been used for feature extraction in line defect detection [1]. Texture analysis has shown great potential in detecting defects under laboratory conditions [10], [13], [28], [40]. Impulse-response testing and statistical pattern recognition are also employed by researchers, as observed in a recent study on defect recognition in concrete plates [26].

However, despite the popularity and effectiveness of these traditional methods in controlled environments, they struggle to perform optimally under challenging conditions, such as poor and complex lighting conditions. This makes defect detection challenging, especially for small, random, and/or unexpectedly located defects. The limitations of conventional machine vision models in detecting defects in challenging settings highlight the importance of exploring alternative methods.

### B. Machine Learning Approaches

Machine learning models have the potential to significantly enhance defect detection in various industrial settings, allowing for accurate and efficient defect identification without direct interaction with technology. This section reviews the most relevant machine learning approaches for defect detection, including deep learning, which provides more precise outcomes than traditional statistical methods, especially in complex production environments. These advanced methods can effectively detect defects that may be difficult to identify using conventional methods, leading to significant cost savings and improved productivity. There is an unequivocal need for general solutions to solve complex challenges that can be served by deep learning [44].

For instance, Zhang et al. [37] employed the attention graph convolution network (AGCN) to effectively detect surface defects in steel plates. Shi et al. [27] utilized YOLOv5 with an attention mechanism, while Boikov et al. [7] employed U-net for the same purpose in steel plates. In the domain of subway tunnels defect detection, an improved Unet approach was proposed by researchers [34]. Road defects were addressed using autoencoders [29], chip defects were detected using YOLOv3 [36], and concrete crack detection was achieved using a Convolutional Neural Network (CNN) [30]. Defect detection in apples was accomplished by researchers [38], and manufacturing defects were targeted using DL-based techniques [35]. A comprehensive overview of DL-based defect detection techniques across various industries can be found in [31], [39], [43], [44], [45].

To tackle the challenges presented in real-world environments, ongoing research efforts are focused on enhancing these DL-based models' accuracy, efficiency, and robustness. Notably, image processing techniques have been employed to improve the performance of DL models. For instance, Haffner et al. [11] used entropy to enhance the accuracy of CNNs in weld segmentation. Data augmentation requirements have also been addressed through techniques such as affine and color transformations, as demonstrated by Nagaraju et al. [20]. CPU-GPU parallel algorithms have also been leveraged to expedite defect segmentation in quality inspection processes [17]. Feature extraction and selection have been improved using the lion optimization algorithm (LOA) for crack detection with CNNs [32]. Furthermore, adopting the Human Visual System (HVS) has been explored to generate an image quality index for defect detection [46].

However, due to ambiguity in inspector labeling, these methods also struggle to adapt to varied environments and requirements, especially in poor or complex lighting conditions. These limitations pose a challenge in the practical application of defect detection methods.

## III. PROPOSED METHOD

To detect defects in complex industrial settings with challenging lighting conditions, we propose using InspectNet. This hybrid deep learning method combines a neural network with traditional Image Processing Techniques (IPT) to accurately and efficiently detect flaws. Our approach leverages the strengths of IPT for image enhancement and deep learning methods for defect detection automation. It applies IPT to pre-process the input image and improve image quality, resulting in a strong and effective solution for surface defect detection. The proposed InspectNet method is structured as depicted in Fig. 2.

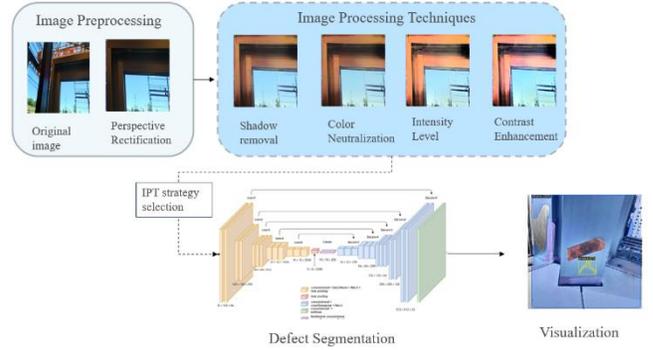

**Fig. 2.** Proposed Methodology of Window Frames Defect Detection.

### A. Data Collection

Accurate data collection plays a pivotal role in the success of any machine learning-based defect detection system. In this study, we meticulously collected real-world data using two primary sources. Firstly, we utilized a cellphone camera provided by a company to capture images of window frames, both with and without defects. Additionally, we employed the Spot Robot equipped with a Pan-Tilt- Zoom (PTZ) camera to further expand our dataset.

During the data collection process, we ensured a diverse range of conditions by varying the lighting, angles, and zoom levels. This encompassed capturing images of different objects, including the outside environment, to guarantee the proposed algorithm's robust performance across various scenarios, as elucidated in Section IV. To ensure the accuracy of our dataset, we manually annotated the collected images, precisely marking the locations of the window frames and the detected defects. The dataset was annotated with four distinct classes: window frames, dents, bends, and scratches. The defect samples were collected from both aluminum window frame pieces and wholes, covering multiple scenarios for a comprehensive evaluation of the proposed algorithm. Fig. 3 shows examples of different window frame defects collected.

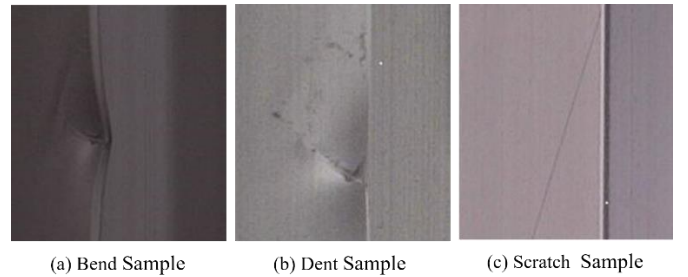

(a) Bend Sample    (b) Dent Sample    (c) Scratch Sample

**Fig. 3.** Types of defects in the collected data set.

### B. Perspective Correction (PR)

Before implementing our method in the cellphone dataset, we pre-processed this data. We corrected the perspective of the images by applying manual simple homography transformation, which improved the view of the window frame and enabled better defect detection as shown in Fig. 4.

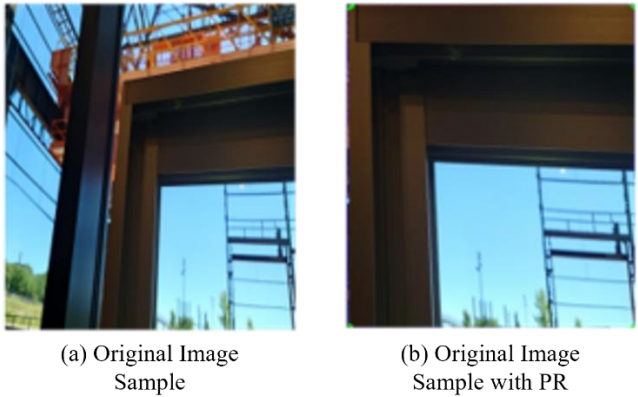

(a) Original Image Sample    (b) Original Image Sample with PR

**Fig. 4.** Perspective Rectification Example.

*C. Image Processing Techniques*

This subsection details the Image Processing Techniques (IPT) employed to pre-process the images, preparing them for input into the Neural Network. The datasets used in our experiments include images captured under various lighting conditions and with various material colors, which could impact the image quality and, consequently, the performance of our model. To address this challenge, we applied several IPTs:

1) *Shadow Removal (SR):* To remove any shadows present in the images that could negatively impact image quality.
2) *Color Neutralization (CN):* To balance and homogenize images' color distribution.
3) *Intensity Level Neutralization (IN):* Correct for images' brightness level variations.
4) *Contrast Enhancement (CE):* Correct for image contrast level variations, facilitating interpretation and making the images more distinguishable.

### 1) Shadow Removal Process (SR)

The shadow removal process is performed using a pretrained Dual Hierarchical Aggregation network (DHAN) proposed by Cun et al. [8]. The DHAN network is based on the VGG16 network, a wellknown convolutional neural network architecture, and uses the context aggregation network (CAN) as the encoder.

The DHAN network uses a series of dilation convolutions and hierarchical aggregation of multicontext features to identify and remove shadows from the images. The network has been pretrained with weights provided by the authors, and we use the network inference procedure to remove shadows from our complete data set (as shown in Fig. 5).

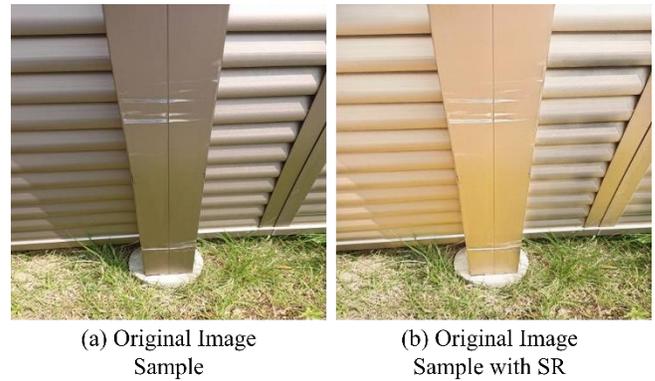

(a) Original Image Sample    (b) Original Image Sample with SR

**Fig. 5.** Comparative sample using the Shadow Removal technique.

### 2) Color Neutralization (CN)

To improve color definition in the data set after shadow removal, we employed a chromatic adaptation transformation known as the Von Kries chromatic adaptation transformation [16]. The Von Kries chromatic adaptation transforms from the source to the target color in the LMS (Long, Medium, Short) color space. This transformation aims to adapt the RGB illuminant color of the data set samples to different illuminates, thus maintaining the constant color white. This results in greater color constancy and improved feature extraction (see Fig. 6).

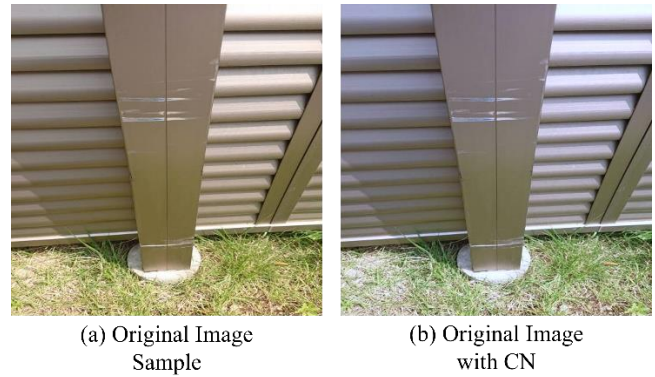

(a) Original Image Sample    (b) Original Image with CN

**Fig. 6.** Comparative sample using the Color Neutralization technique.

### 3) Intensity Level Neutralization (IN)

Multi-Scale Retinex (MSR) was applied on the intensity channel. The output colors are computed so that chromaticity remains the same as in the original image. This filter helps preserve this relative lightness, ensuring that image intensity levels are neutralized without altering the chromaticity and color composition. This process helps remove the intensity variations that could affect feature extraction and further image data processing (see Fig. 7).

This algorithm is derived from Land and Maccan [15], who established that the visual system does not perceive absolute lightness but rather relative lightness; i.e., lightness variations in local image regions.

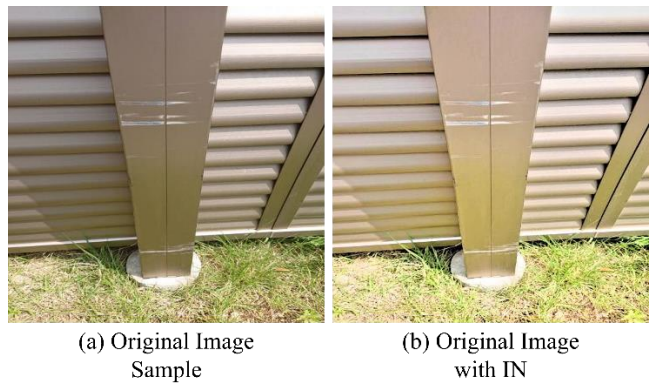

**Fig. 7.** Comparative sample using the Intensity Neutralization technique.

#### 4) Contrast Enhancement (CE)

The final step in the image processing technique is enhancing the contrast. We transform the data set samples into RGB channels using a histogram equalization method [19]. This process helps achieve a better visual representation of the image, facilitating the detection of defects and other anomalies. By equalizing the histogram, overall image contrast is increased, leading to improved feature representation and better results in the deep-learning pipeline (see Fig. 8).

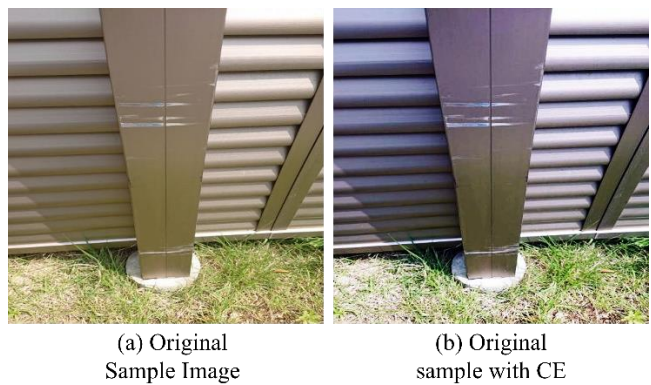

**Fig. 8.** Comparative sample using the Contrast Enhancement technique.

### D. Machine Learning Model

This subsection introduces the core component of InspectNet: a neural network architecture specifically designed for accurate defect detection. The InspectNet network is a hybrid model that incorporates two key components: the Resnet152 [12] and U-net neural network [25]. Transfer learning is utilized, where the Resnet network serves as the Encoder. In this architecture, the last fully connected layer of the Resnet network is discarded. Instead, it is integrated with a Decoder that expands the feature map obtained by the Encoder and generates a mask. This combination enables the network to effectively capture and localize defects. The net- work takes pre-processed images as input, which have undergone the Image Processing Technique (IPT) strategy and have been resized to a resolution of 500x500 pixels. Additionally, ground truth masks, serving as references for the network, are also provided as input. These masks provide the network with essential information for learning and accurately detecting defects in the input images.

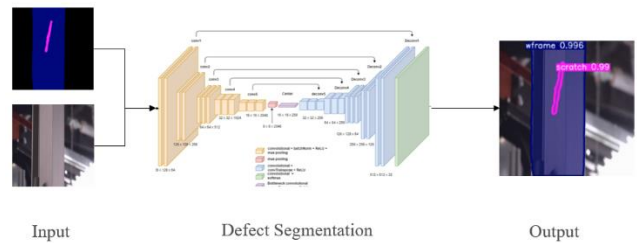

**Fig. 9.** Visualization of the Deep Neural Network.

### IV. EXPERIMENTS AND RESULTS

In this section, we detail our experimental methodology and results aimed at discovering the optimal Image Processing Technique (IPT) strategy to enhance deep learning model accuracy when applied to robotic images of window frames. Specifically, we use a tool to compare the performance of our neural network with and without applying different IPT strategies to identify the best approach. We then quantitatively evaluate the impact of this optimal IPT strategy on the performance of our model for detecting window frame defects. Finally, we compare the results of our proposed method with those of other state-of-the-art techniques, providing insight into our contribution to the field.

### A. Data Collection

The performance evaluation for the proposed model involved utilizing three primary datasets. The first dataset was created by capturing images of multiple window frames using various cellphone cameras. Two datasets were obtained from the SPOT datasets' PTZ camera as well, with each collected in separate laboratory settings. The first lab dataset comprised multiple window frame pieces under diverse conditions, while the second lab dataset consisted of a single complete window frame that was captured across different views using the PTZ camera. These datasets provided a comprehensive range of scenarios for evaluating model effectiveness. The dataset employed in the experiments contained window frames exhibiting three types of defects: dents, bends, and scratches. The images were captured under varying lighting conditions, orientations, and zoom levels. Further details about the dataset are provided below. All experiments were conducted on a GPU, utilizing the PyTorch framework to efficiently process and analyze the collected data.

#### 1) Dataset by Cellphones (Cell)

This dataset comprises 441 high-resolution RGB images of five window frame objects, both inside and outside. It provides a comprehensive set of images of window frames captured under various lighting conditions and distances by a simple cellphone camera. This data set is valuable in evaluating the robustness of our proposed method against external lighting condition variations. An example of this data set appears in Fig. 10.

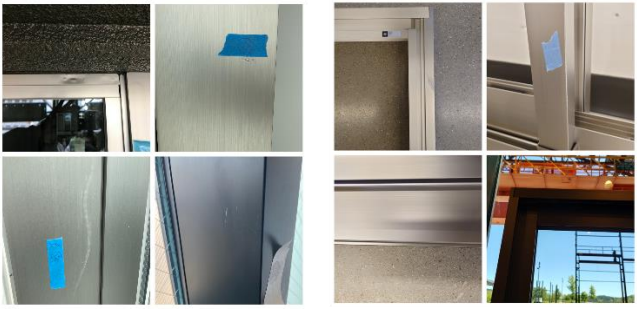

**Fig. 10.** Dataset Examples from Cellphone Camera.

*2) Dataset by Spot Robot (SPOT-Lab 1)*

The SPOT-Lab 1 Dataset includes 200 color images of ten different window frame pieces captured under laboratory conditions. The Spot in the Lab data set was gathered using the Spot robot. This set contained images captured from 0,15,30,45, and 60 degrees of inclination, with variations in four color material samples (YS-1N (Silver), YH-1N (Stainless), YB-5N (Bronze), YK-1N (Black)), two illuminance conditions (normal ambient light and using Spotlight), and zoom conditions (0, 5X, 10X, and 30X). This dataset is valuable to evaluate the performance of our proposed method under controlled conditions.

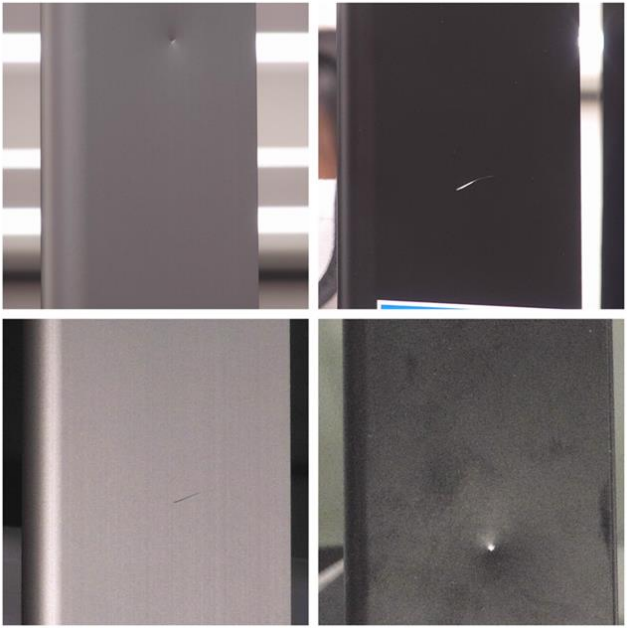

**Fig. 11.** Dataset by PTZ camera obtained by the Spot robot in Lab 1.

*3) Dataset by Spot Robot (SPOT-Lab 2)*

The SPOT-Lab 2 Dataset includes 80 color images of one specific window frame captured under stable lighting conditions in a cluttered environment. This dataset was gathered using the Spot robot and contained images captured from three different view angles (center, right and left) with no variations in color samples, lighting conditions, or zoom conditions. This dataset is valuable in evaluating the performance of our proposed method under controlled conditions.

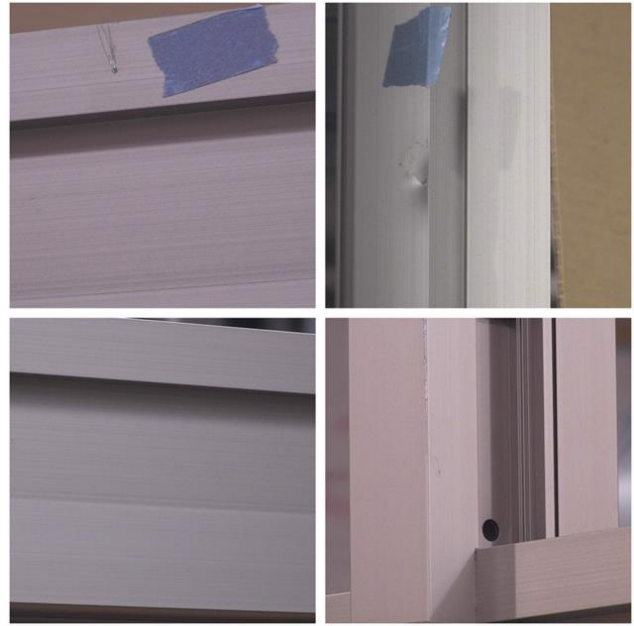

**Fig. 12.** Dataset by PTZ camera obtained by the Spot robot in Lab 2.B.

*B. Data Labelling*

To start, we used the Roboflow platform [18] to accurately label the images for four defects: bend, scratch, and dent. These class labels were then used to generate the necessary masks for the training process of our network. The annotation process produced a set of COCO format images, each with a size of 500 x 500 pixels. We later used these images to create the ground truth masks, which were used as a reference for evaluating the effectiveness of our proposed method.

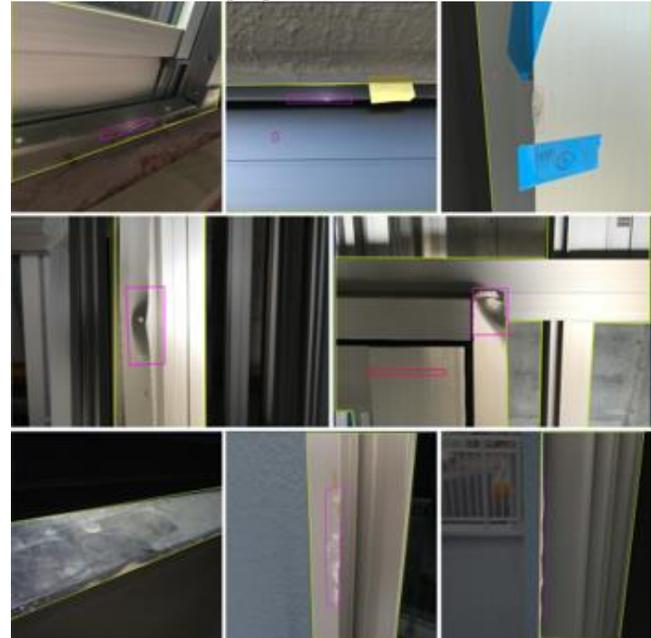

**Fig. 13.** Labelling Process Example.

*C. IPT Implementation Process*

We conducted an exhaustive analysis of all possible parameter combinations to determine our datasets' most effective combination of IPT parameters.

We subsequently partitioned the dataset into 16 subsets, each representing a different IPT strategy, to evaluate the impact of each method separately using the same dataset. The next step was to assess the performance of each dataset using our NN model. The images below demonstrate that defects have become apparent through our IPT strategies. The preliminary results appear in the following figures:

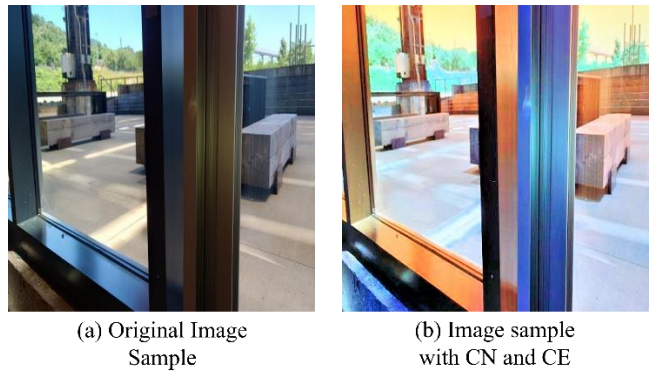

**Fig. 14.** Comparative sample after Color Neutralization and Contrast Enhancement.

*D. IPT Impact Analysis*

We tested our model, InspectNet, by training it on each dataset generated by applying different IPT combinations to the original dataset. Following a thorough evaluation, we saved the best-performing model and recorded its corresponding loss. This performance metric measures the degree of overlap between predicted and ground-truth object segmentation masks, indicating the accuracy and robustness of the algorithm in identifying and localizing objects in the image. The end goal of this process is to provide the Neural Network with a robust and reliable input leading to improved defect detection performance. This integration process is crucial since it directly affects the quality of the final results, and the ability of the network to learn and recognize relevant features from the images.

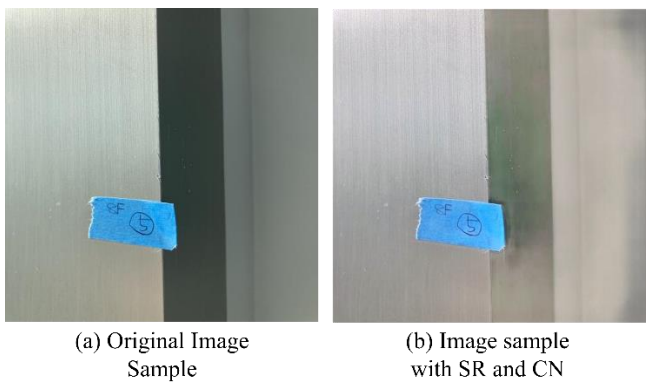

**Fig. 15.** Comparative sample after SR and CN.

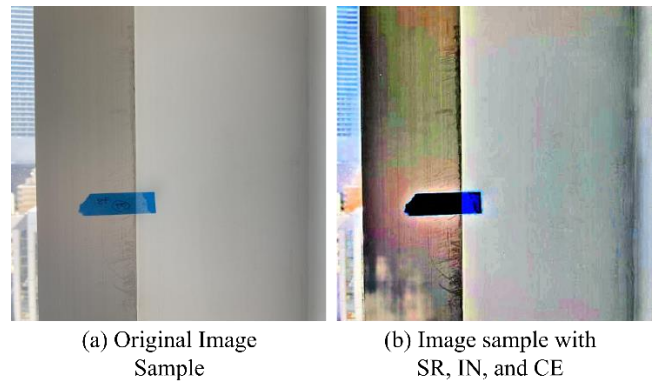

**Fig. 16.** Comparative sample after Shadow Removal, Intensity Neutralization, and Contrast Enhancement techniques.

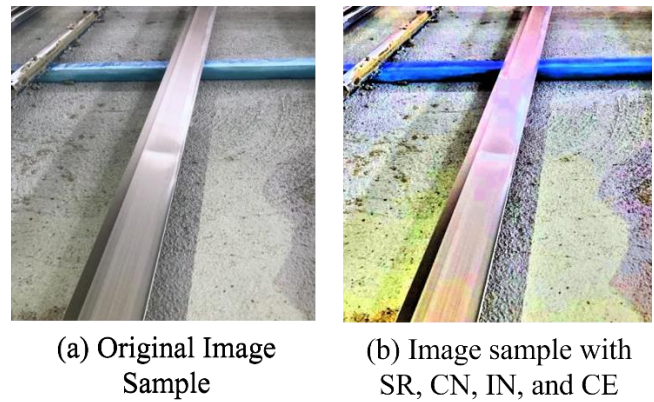

**Fig. 17.** Comparative Sample after Shadow Removal, Color Neutralization, Intensity Neutralization, and Contrast Enhancement Techniques.

Our data interpretation highlights that Contrast Enhancement emerged as the most effective Image Processing Technique (IPT) for enhancing the accuracy of detecting scratch defects. Interestingly, we observed that more than half of the strategies employed actually decreased the accuracy. This suggests that, in the case of scratch defects, the application of shadow removal adversely affected the accuracy of detection. These results are presented in Fig. 18 and 19.

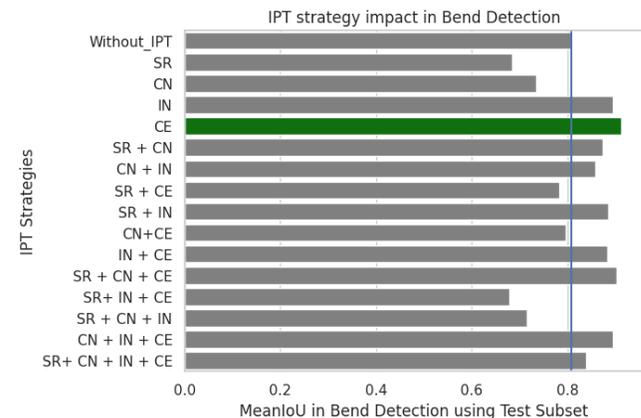

**Fig. 18.** IPT Strategies' Impact in Bend Detection using InspectNet.

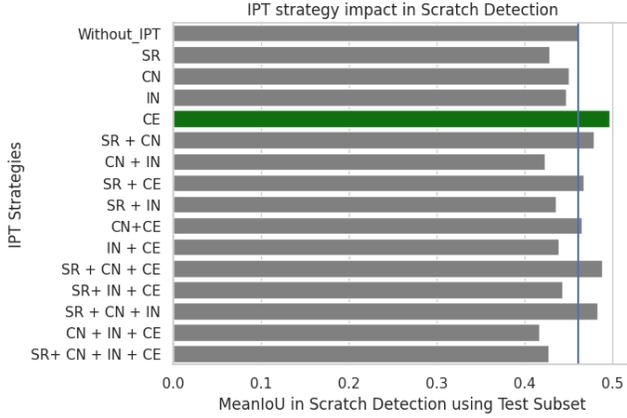

**Fig. 19.** IPT strategies' Impact in Scratch Detection using InspectNet.

Furthermore, upon visualization, it became apparent that scratch defects present a greater challenge in terms of detection. This was evident from the fact that the best Intersection over Union (IoU) score achieved for scratch defects was consistently below 0.5. The lower IoU scores indicate that accurately capturing and delineating scratch defects proved to be more difficult compared to other types of defects. This is shown in Fig. 19.

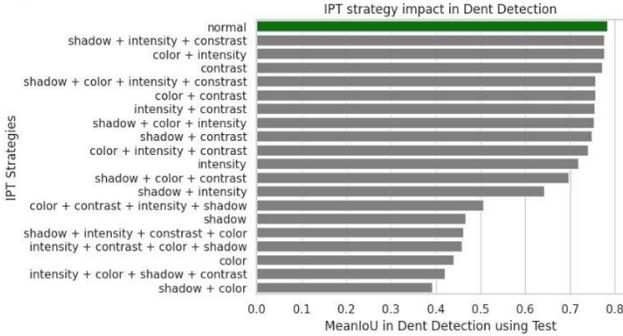

**Fig. 20.** IPT Strategies' Impact in Dent Detection using InspectNet.

Our objective was also to determine whether the order of operations significantly affected the results and to identify the most appropriate technique for detecting specific defects. We found that IPT order was irrelevant to detect the window frame, as shown in Fig. 21.

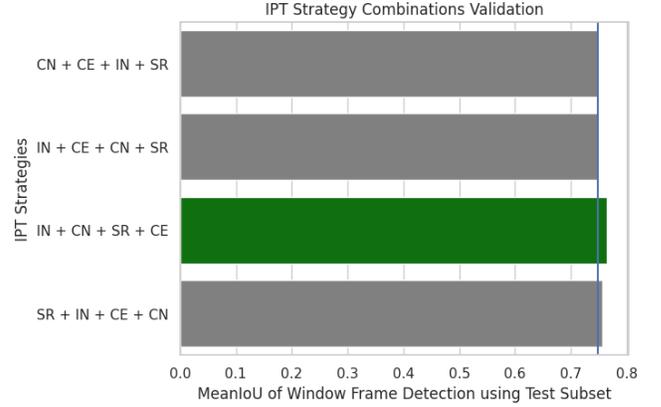

**Fig. 21.** Comparison of different IPT Strategies Combination in window frames detection.

*E. Benchmarking Analysis*

To validate the effectiveness of our hybrid model, we conducted experiments comparing it to other machine-learning pipelines. We used the non-rectified Unet model in order to conduct separate analyses on a dataset of 500 samples for the Training subset, 200 for the Valid subset, and 55 for the Test subset. We compared the defect detection accuracy of our proposed method to Unet model using the meanIoU metric in the Test subset. Through this series of comparative experiments, we demonstrated the superior performance of our proposed method. The results of these experiments appear in Table 2 and provide a comprehensive evaluation of how effective our method is, compared to other techniques.

TABLE I
BENCHMARKING ANALYSIS TABLE

| | Test Subset | | |
|---|---|---|---|
| **Defect Class:** | **Bend** | **Dent** | **Scratch** |
| meanIoU Unet | 0.80 | 0.76 | 0.46 |
| **meanIoU InspectNet (Ours)** | 0.91 | 0.81 | 0.50 |
| **Improvement** | 13.75% | 6.58% | 8.67% |

V. DISCUSSION

Our study aimed to improve defect detection ac- curacy in complex and dynamic environments by incorporating Image Processing Techniques (IPT) into a deep learning pipeline. We proposed a hybrid method, InspectNet, that combines a U-net model with best strategy IPTs, and compared its performance to a baseline deep learning model without IPT. Our experiments also revealed that perspective rectification of the cellphone dataset improved our results, as shown in Fig. 4.

Our results showed a significant increase in defect detection accuracy when using InspectNet compared to the baseline model. We found that IPT order was irrelevant to detect window frame, as seen in Fig. 21.

However, we also observed that Shadow Removal using GAN actually harmed the accuracy of detecting defects. This is likely because the generated image may not accurately reflect the original, ultimately hindering the performance of U- net. Notably, our model demonstrated superior performance, particularly when working with smaller datasets as it is shown in Fig 3a. This is significant because it suggests that InspectNet may be a useful tool to detect defects in scenarios where data collection is challenging. In practical terms, our findings could have important implications for industries that rely on defect detection, such as manufacturing or quality control. This impact is visualized in Fig. 22.

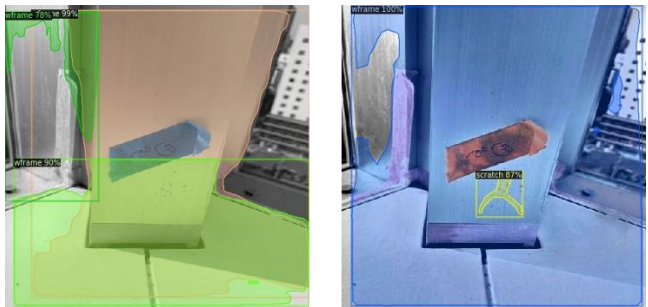

(a) Results in normal Unet  (b) Example of using best IPT strategy

**Fig. 22.** Comparative Sample after Shadow Removal, Color Neutralization, Intensity Neutralization, and Contrast Enhancement Techniques.

## VI. CONCLUSIONS

Based on the results of our experiments, we conclude that the proposed method, InspectNet, is a practical hybrid approach for defect detection in complex lighting conditions and environments. Our findings show that using a specific IPT strategy as a pre-processing step significantly improves the performance of our deep-learning pipeline. This study demonstrates the potential of integrating defect detection, providing a new approach to quality control in various industrial settings. Our hybrid method demonstrates exceptional results when applied to RGB images, achieving a high accuracy that outperforms traditional deep-learning methods. While the proposed method shows promising results, further research is necessary to optimize the approach for more defects and enhance its robustness to other delicate objects. With such high precision, our method is well-suited for many applications, including object recognition and medical imaging. To enhance the self-adaptation capabilities of InspectNet, several areas of exploration for future research exist. Incorporating advanced pose and lighting estimation techniques in specific scenarios has emerged as a promising avenue for enhancing the accuracy and efficiency of the inspection process. To further explore this approach, it is crucial that we expand the scope of our analysis beyond the defects, pose angles, and lighting issues addressed in this paper. We aim to create an autonomous inspection system that operates efficiently and adaptively in complex environments, delivering high-quality results. Continued research and development will enable us to integrate advanced techniques such as optimizing viewpoint planning and accounting for lighting conditions, further improving the performance of the system.

TABLE II
DIFFERENT IPT STRATEGIES RESULTS

| IPT Strategy | Test Bend IoU | Test Dent IoU | Test Scratch IoU | Test Wframe IoU |
|---|---|---|---|---|
| Without IPT | 0.808 | 0.764 | 0.461 | 0.746 |
| SR | 0.685 | 0.731 | 0.428 | 0.748 |
| CN | 0.735 | 0.765 | 0.451 | 0.754 |
| IN | 0.896 | 0.785 | 0.447 | 0.756 |
| CE | 0.913 | 0.748 | 0.497 | 0.767 |
| SR + CN | 0.874 | 0.740 | 0.479 | 0.784 |
| CN + IN | 0.858 | 0.808 | 0.423 | 0.742 |
| SR + CE | 0.783 | 0.792 | 0.467 | 0.755 |
| CN + CE | 0.797 | 0.757 | 0.465 | 0.767 |
| IN + CE | 0.884 | 0.743 | 0.439 | 0.754 |
| SR + CN + CE | 0.903 | 0.762 | 0.488 | 0.756 |
| SR + IN + CE | 0.678 | 0.761 | 0.443 | 0.710 |
| SR + CN + IN | 0.715 | 0.771 | 0.483 | 0.759 |
| CN + IN + CE | 0.894 | 0.777 | 0.417 | 0.739 |
| SR + CN + IN + CE | 0.840 | 0.789 | 0.427 | 0.736 |

TABLE III
DIFFERENT IPT STRATEGY COMBINATION RESULTS

| Test Metrics | SR + IN + CE + CN | CN + CE + IN + SR | IN + CN + SR + CE | IN + CE + CN + SR |
|---|---|---|---|---|
| Bend IoU | 0.841 | 0.825 | 0.811 | 0.841 |
| Dent IoU | 0.790 | 0.716 | 0.771 | 0.737 |
| Scratch IoU | 0.479 | 0.441 | 0.428 | 0.472 |
| Window frame IoU | 0.755 | 0.749 | 0.765 | 0.747 |

### DECLARATIONS

**Conflict of interest:** the authors declare that they have no known competing financial interests or personal relationships that could have appeared to influence the work reported in this paper.

### APPENDIX

Our different combinations of IPT strategies impact results are shown in the tables II and III.

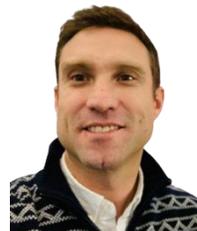

**Jorge Vásquez** (Talcahuano, Chile, June 12nd 1982) is Ph.D. student at Carnegie Mellon University. He received an M.S. degree in Mechanical Engineering and an M.S. degree in Computational Design both from Carnegie Mellon University, BS In Mechanical Engineering from the University of Santiago, and a BS in Systems Engineering from the Chilean Army. His current research work is concerned with image processing techniques to improve deep learning models for defect detection. He is part of the Computational Engineering and Robotics Lab (CERLAB) at Carnegie Mellon University.

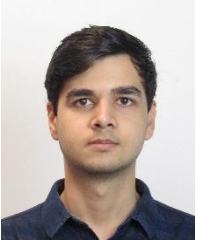
**Hemant K. Sharma** received a Master of Science in Mechanical Engineering - Research degree with a strong computer vision and machine learning background from Carnegie Mellon University in 2023. He will join Triangulate Labs as Senior Computer Vision Engineer working on skin cancer detection.

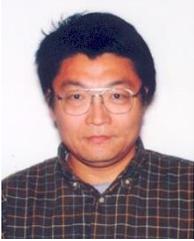
**Tomotake Furuhata** received his M.S. in Mechanical Engineering from the University of Tokyo. He is currently a Research Scientist in the Computational Engineering and Robotics Lab (CERLAB) of the Department of Mechanical Engineering at Carnegie Mellon University. His research topics are computational geometry, computer vision, image processing techniques, machine learning, and robotics for industrial applications.

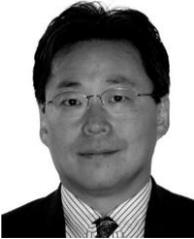
**Kenji Shimada** received the Ph.D. degree from the Massachusetts Institute of Technology in 1993. He received his Ph.D. from Massachusetts Institute of Technology in 1993, after which he returned to IBM Japan, commercialized BubbleMesh, which he invented in his Ph.D. thesis, and managed the Advanced Computer Graphics Group. He moved to the United States in 1996 to join the faculty of Carnegie Mellon University and served as an Assistant Professor, an Associate Professor, and a Professor in the Department of Mechanical Engineering, the Robotics Institute, the Department of Biomedical Engineering, and the Department of Civil and Environmental Engineering. He is currently a Theodore Ahrens Professor of engineering with Carnegie Mellon University.